\documentclass[%
superscriptaddress,
amsmath,amssymb,
 aps,
]{revtex4-2}

\usepackage{graphicx}
\usepackage{dcolumn}
\usepackage{bm}
\usepackage{setspace}
\usepackage{hyperref}

\begin{document}

\preprint{ArXiv/manuscript}

\title{Efficient Active Flow Control Strategy for Confined Square Cylinder Wake Using Deep Learning-Based Surrogate Model and Reinforcement Learning}

\author{Meng Zhang}
\affiliation{School of Mechanical Engineering, Pusan National University, 2, Busandaehak-ro 63beon-gil, Geumjeong-gu, Busan, 46241, Republic of Korea}

\author{Mustafa Zuhair Gheni Yousif}%
\author{Minze Xu}
\author{Haifeng Zhou}
\author{Linqi Yu}
\author{Hee-Chang Lim}

\email[]{Corresponding author, hclim@pusan.ac.kr}

\affiliation{School of Mechanical Engineering, Pusan National University, 2, Busandaehak-ro 63beon-gil, Geumjeong-gu, Busan, 46241, Republic of Korea}

\date{\today}

\begin{abstract}
This study introduces a deep learning surrogate model-based reinforcement learning (DL–MBRL) for the active control of two-dimensional (2D) wake flow past a square cylinder confined between parallel no-slip walls using antiphase jets. In this DL–MBRL process, a deep neural network agent alternates between interacting with a deep learning-based surrogate model (DL–SM) and a computational fluid dynamics (CFD) simulation to suppress wake vortex shedding, thereby significantly reducing computational costs. The DL–SM, which incorporates a Transformer and a multiscale enhanced super-resolution generative adversarial network (MS–ESRGAN), effectively models complex nonlinear flow dynamics and emulates the CFD environment efficiently. Trained on 2D direct numerical simulation (DNS) wake flow data, the Transformer and MS–ESRGAN showed excellent agreement with DNS results, validating the accuracy of the DL–SM. Error analysis indicated the need to replace the DL–SM with the CFD environment every five interactions to maintain reliability. Although DL–MBRL exhibited less robust convergence than model-free reinforcement learning (MFRL) in the training process, it achieved a 49.2\% reduction in training time, decreasing from 41.87 h to 20.62 h. Both MFRL and DL–MBRL demonstrated approximately a 98\% reduction in shedding energy and a 95\% reduction in the standard deviation of $C_L$ during testing. However, MFRL exhibited a nonzero mean lift coefficient owing to insufficient exploration. In contrast, DL–MBRL leveraged the inherent randomness of the DL–SM to enhance exploration, effectively addressing the nonzero mean $C_L$ issue. Comprehensive qualitative and quantitative assessments demonstrated that DL–MBRL is not only comparably effective but also superior to MFRL in flow stabilisation, with significantly reduced training time. This study underscores the potential of combining Deep Reinforcement Learning with DL–SM to improve the efficiency and accuracy of active flow control.
\end{abstract}

\maketitle

\section{Introduction}\label{sec1}
Suppressing vortex shedding behind bluff bodies is crucial in aerodynamics and hydrodynamics applications because it significantly decreases unsteady forces on structures, reduces structural vibrations, and enhances flow efficiency \citep{Rashidietal2016, Bovandetal2015, Zangetal2013}. Flow control techniques are pivotal across diverse fields, such as aerospace engineering, automotive design, and energy systems. They effectively manage turbulence, enhance aerodynamic stability, minimise drag, and optimise energy usage. Flow control techniques can be broadly categorised into passive flow control and active flow control (AFC). Passive flow control \citep{Joshi&Gujarathi2016} manipulates fluid flow using fixed features without external energy. Examples include animal-inspired designs such as shark skin surfaces \citep{Dean&Bhushan2010}, fish scale bionics \citep{Wuetal2018}, and other methods such as vortex generators \citep{Fouatihetal2016}, dimples \citep{Stanlyetal2016}, and ribs \citep{Leeetal2005}. Unlike passive flow control, which relies on fixed geometrical features, AFC techniques use external energy sources or actuators to intentionally manipulate fluid flow \citep{Joshi&Gujarathi2016}. Various AFC methods have been developed, including synthetic jets \citep{Wangetal2016, Li&Zhang2022, Rabaultetal2019}, plasma actuators \citep{Jukes&Choi2009, Thomasetal2008, Yousifetal2023a, Yousifetal2024}, and cylinder rotation \citep{Muddada&Patnaik2010, Dipankaretal2007, Zhuetal2015}.

Owing to advancements in big data, computing power, and algorithm development, machine learning (ML) has attracted extensive attention in recent decades and has been applied across various fields, such as computer vision \citep{Zhangetal2018}, natural language translation \citep{Collobertetal2011}, and autonomous driving \citep{Grigorescuetal2020} and so on. In fluid dynamics, ML has been used to address several challenges, including flow denoising and reconstruction \citep{Fukamietal2019, Liuetal2020, Yousifetal2021, Yousifetal2023b, Yousifetal2022a, Zhangetal2023}, flow prediction \citep{Lee&You2019}, turbulent inflow generation \citep{Yousifetal2022b, Yousifetal2023c}, and AFC \citep{Rabaultetal2019, Li&Zhang2022}. Deep Reinforcement Learning (DRL), which is a ML method where an agent learns to make decisions by interacting with an environment, has shown remarkable results in fields such as robotics \citep{Koberetal2013}, game-playing \citep{Silveretal2017, Silveretal2018}, and optimisation problems \citep{Belloetal2016, Mazyavkinaetal2021}. DRL also significantly contributes to solving high-dimensional, nonlinear problems in fluid dynamics, particularly for flow control \citep{Rabaultetal2019, Li&Zhang2022, Yousifetal2023a}, shape optimisation in the aerospace and automotive industries \citep{Matosetal2004, Muyletal2004, Lamptonetal2008}, and process optimisation \citep{Leeetal2018, Novatietal2016}. This research aims to enhance the utilisation of DRL in flow control to suppress vortex shedding and improve flow stability.

Incorporating DRL into fluid dynamics experiments faces challenges such as sensing and actuation limitations, signal noise, and data transmission delays \citep{Fanetal2020}. Consequently, current research that combines DRL with AFC techniques primarily relies on computational fluid dynamics (CFD) for precise, fast, and cost-effective simulations. Rabault et al. \citep{Rabaultetal2019} pioneered the application of DRL to CFD simulations by conducting AFC in a two-dimensional (2D) simulation of the Kármán vortex street with two jets on the sides of a cylinder. Their work resulted in the successful stabilisation of the vortex alley and an approximate 8\% reduction in drag on the flow at a moderate Reynolds number. Building on this work, Rabault and Kuhnle \citep{Rabault&Kuhnle2019} accelerated the DRL training process by parallelising it across multiple environments, which enabled faster experience collection and increased execution speed. Tang et al. \citep{Tangetal2020} introduced a smoothing interpolation function to allow DRL to set continuous actions for controlling four synthetic jets around a cylinder in a 2D flow domain. Their findings showed that DRL effectively reduced lift and drag fluctuations at both trained and untrained Reynolds numbers, demonstrating its interpolation and extrapolation capabilities. Yousif et al. \citep{Yousifetal2023a} employed a DRL-based technique to control flow around a square cylinder by adjusting the AC voltage amplitude across three distinct configurations of multiple plasma actuators. Li and Zhang \citep{Li&Zhang2022} enhanced the efficacy of DRL-based flow control for suppressing vortex shedding around a cylinder confined between two walls. They integrated physical insights from global stability and sensitivity analyses, which was particularly beneficial owing to strategically placed probes in sensitive regions.

The abovementioned studies on DRL-based AFC underscore its capability to surpass traditional methods, particularly in intricate nonlinear flow scenarios. This advantage stems from DRL's inherent data-driven nature, enabling the exploration of new control strategies without dependence on analytical derivations. However, this characteristic also necessitates numerous interactions with the environment, requiring hundreds of iterations even for basic flow control problems to achieve convergence \citep{Weiner&Geise2024}. This requirement results in high computational costs and substantial time consumption, constraining its applicability for dynamic control \citep{Liu&Wang2021, Linotetal2023}. Researchers have proposed various strategies to alleviate this computational burden, including the use of surrogate models to emulate the real environment in DRL frameworks. The surrogate model, often referred to as a method utilising various established simple function approximators (e.g. Gaussian processes, linear models, and Gaussian mixture models \citep{Deisenroth&Rasmussen2011, Tassaetal2012, Levine&Koltun2013}) or complex function approximators based on deep learning, aims to emulate the real environment. This approach leverages surrogate models for inference, effectively reducing the computational costs associated with real-world environments, a methodology known as model-based reinforcement learning (MBRL). Liu and Wang \citep{Liu&Wang2021} developed a physics-informed MBRL framework that integrates prior knowledge of the environment's physics, enhancing the surrogate model's accuracy and minimising the requirement for interactions. They applied this framework to classic control problems such as cart–pole dynamics, pendulum systems, and simple one-dimensional turbulent dynamics. Zeng et al. \citep{Zengetal2022} integrated an autoencoder with neural ordinary differential equations (Neural ODEs) to develop a low-dimensional surrogate model for DRL training. This approach effectively stabilised a forced equilibrium solution of the Kuramoto–Sivashinsky equation system. Building on this work, Linot et al. \citep{Linotetal2023} further enhanced the framework for drag reduction in plane Couette flow, employing two slot jets on one wall.

To our knowledge, only the three abovementioned studies have used MBRL for AFC, and their methodologies possess inherent limitations. First, the surrogate models proposed above require complete flow field data for interaction with the DRL agent in each cycle. These flow fields are highly detailed, resulting in a high-dimensional state space when directly input to the DRL agent, thereby complicating training. Second, an effective surrogate model in DRL for AFC must precisely replicate the real environment, encompassing temporal dependencies, capturing long-term memory, modelling complex nonlinear dynamics, and generalising to future data. However, the abovementioned surrogate models, incorporating Neural ODEs and long short-term memory (LSTM) \citep{Hochreiter&Schmidhuber1997}, are limited in their capabilities. Lastly, while the proposed MBRL approaches have shown effectiveness in addressing basic dynamic control tasks for initial validation, additional research is necessary to effectively address complex real-world challenges.

This study introduces a deep learning surrogate model-based DRL (DL–MBRL) approach for actively controlling the flow past a square cylinder confined between parallel no-slip walls. DL–MBRL operates in two alternating environments: a deep-learning-based surrogate model (DL–SM) and a CFD simulation. By leveraging the DL–SM for efficient approximation and the CFD simulation for high accuracy, this approach aims to achieve precise flow control, particularly in suppressing wake vortex shedding, while significantly reducing computational costs. To emulate the CFD environment effectively, the DL–SM uses the Transformer \citep{Vaswanietal2017}, which is known for its superior parallelism and efficient sequence processing enabled by the attention mechanism. Furthermore, the DL–SM integrates a multiscale enhanced super-resolution generative adversarial network (MS–ESRGAN) \citep{Yousifetal2021} to reconstruct detailed flow fields, ensuring effective collaboration with the CFD environment. The state channel used for interaction between the DL–SM and the DRL agent is characterised by flow field measurements from sparse probes, which reduce dimensionality, streamline the training process, and improve the generalization of the DRL.

The structure of this paper is as follows. Section \ref{sec2} introduces the DNS setup and outlines the typical MFRL framework, followed by the methodology for actively controlling using the proposed DL–MBRL and the detailed DL models in the DL–SM. Section \ref{sec3} elaborates on the convergence of learning, presents the results of the DL–SM, and assesses the control effects achieved by DRL. Finally, Section \ref{sec4} summarises the conclusions drawn from this study.

\section{Methodology}\label{sec2}
This section is structured into four main parts: a methodology for the CFD numerical simulation, a description of the typical DRL framework, an application of MBRL, and details of the deep-learning-based surrogate model. The CFD numerical simulation uses the spectral–element open-source code Nek5000 \citep{Fischeretal2008, Fischeretal2017}. The DRL algorithm is developed using Python 3.6 and the open-source framework Tensorforce 0.4.3 \citep{Kuhnleetal2017}, which is built on the TensorFlow 1.13.1 framework \citep{Abadietal2016}.

\subsection{Direct numerical simulation}\label{subsec1}
In this study, a 2D direct numerical simulation (DNS) of the flow past a square cylinder confined between parallel walls is performed. This involves solving the 2D incompressible Navier–Stokes equations and the continuity equation, which are expressed in a non-dimensional form as follows:
\begin{equation} \label{eqn:eq1}
\frac{\partial {\textbf{\textit{u}}}}{\partial t} + {\textbf{\textit{u}}} \cdot \nabla{\textbf{\textit{u}}} = -\nabla p + \frac{1}{Re} {\nabla}^2 {\textbf{\textit{u}}},
\end{equation}
\begin{equation} \label{eqn:eq2}
\nabla \cdot {\textbf{\textit{u}}} = 0,
\end{equation}
\noindent where $\textbf{\textit{u}}$ denotes the velocity vector, $t$ represents time, and $p$ denotes pressure. All quantities are made dimensionless using the square cylinder width $D$, maximum velocity of the parabolic inflow $u_{max}$, and fluid density $\rho$. The Reynolds number ($Re$) is set to 150, defined as $Re=u_{max}D/\nu$, where $\nu$ is the kinematic viscosity of the fluid. The simulation employs a time step of 0.005 unit time.

\begin{figure}[htbp]
  \centerline{\includegraphics[scale=0.18]{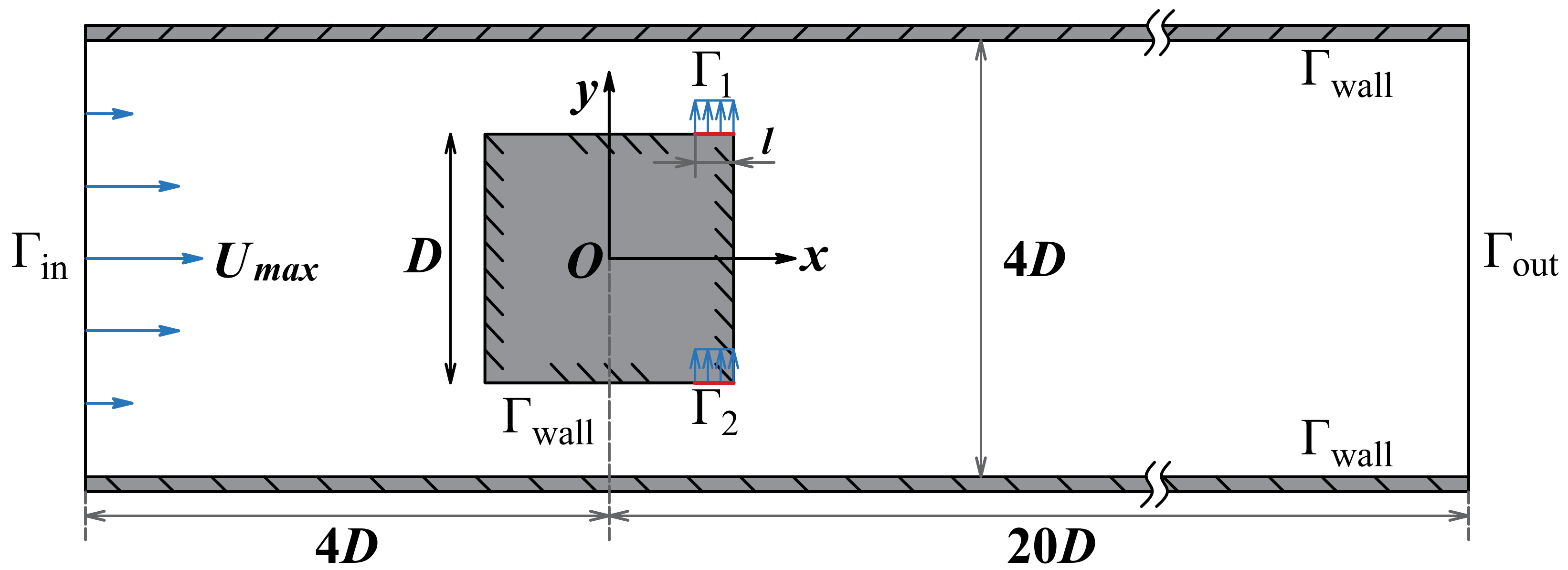}}
  \caption{Geometrical description of the configuration and boundary conditions used for simulating the flow past the square cylinder confined between parallel no-slip walls. The square cylinder, with a width $D$, is equipped with a pair of antiphase jets, each with a width $l $ as indicated by red lines ($\Gamma_{1}$ and $\Gamma_{2}$). These jets feature uniform velocity profiles perpendicular to the cylinder walls. Boundary conditions include $\Gamma_{wall}$ for no-slip conditions on solid walls, $\Gamma_{in}$ for the inflow condition, and $\Gamma_{out}$ for the outflow condition.}
\label{fig:F1}
\end{figure}

Figure~\ref{fig:F1} depicts the configuration and boundary conditions used for simulating the flow past a square cylinder confined between parallel walls. The computational domain is a 2D rectangle with dimensions $L_x \times L_y$ = $24D \times 4D$, where $D$ is the width of the square cylinder. Here, $L_x$ and $L_y$ denote the streamwise and spanwise domain dimensions, respectively. The centre of the square cylinder serves as the origin of the Cartesian coordinate system. The upstream and downstream boundaries are located at distances of $4D$ and $20D$ from the centre of the cylinder, respectively. The surface of the cylinder, as well as the top and bottom walls, adhere to the no-slip boundary condition $\Gamma_{wall}$, where the velocity of the fluid is zero. The inflow condition on the left side, denoted as $\Gamma_{in}$, is specified by a parabolic velocity profile, defined as
\begin{equation} \label{eqn:eq3}
u(y)= (1-y)\times (1+y),
\end{equation}
\begin{equation} \label{eqn:eq4}
v(y) = 0,
\end{equation}
\noindent where ($u(y)$, $v(y) $) represents the non-dimensional velocity vector yielding a maximum inflow velocity of $u_{max}$ = 1. The right side outflow condition $\Gamma_{out}$ is $(p\textbf{\textit{I}} - (1/Re)\nabla \textbf{\textit{u}}) \cdot \textbf{\textit{n}} = 0$, where $\textbf{\textit{n}}$ denotes the outward normal. The 2D flow simulation uses the spectral–element open-source code Nek5000 \citep{Fischeretal2008, Fischeretal2017}. In the spectral–element method (SEM), the computational domain is discretised into multiple elements, and the velocity and pressure fields are approximated using high-order Lagrange interpolants based on Legendre polynomials at the Gauss–Lobatto–Legendre (GLL) quadrature points \citep{Vinuesaetal2018}. In this study, a total of 352 elements are used to discretise the domain with an order $N$ = 7, resulting in 22,528 grid points. The drag coefficient $C_D$, lift coefficient $C_L$, and Strouhal number $St$ are defined as $C_D = F_D/0.5 \rho u_{max}^2 D^2$, $ C_L = F_L/0.5 \rho u_{max}^2 D^2$, and $St = fD/ u_{max}$, respectively. Here, $F_D$ represents the drag force on the cylinder surface, $F_L$ denotes the lift force on the cylinder surface, and $f $ is the vortex shedding frequency. From the simulation, the mean drag coefficient ($\overline{C_D}$) is determined to be 1.60, the range of lift coefficient ($C_{L}^{max}- C_{L}^{min}$) is 0.33, and $St$ is 0.19. These values have been validated against results reported by Turki et al. \citep{Turkietal2003a}, which are $\overline{C_D}$ = 1.62, $C_{L}^{max}- C_{L}^{min}$ = 0.35, and $St$ = 0.18.

As shown in Figure~\ref{fig:F1}, the square cylinder is equipped with a pair of antiphase jets symmetrically positioned on the top and bottom, each with a width $l=2D/25$ and uniform velocity profiles perpendicular to the walls. These jets are employed for AFC, where adjustments in mass flow rates effectively mitigate vortex shedding and stabilise the wake flow behind the cylinder. In this study, the total mass flow rate of the two jets is zero, indicating equal magnitudes with one jet blowing and the other sucking. The parameter $Q$ mentioned in subsequent analyses denotes the magnitude of the jets, defined as follows:
\begin{equation} \label{eqn:eq5}
 Q = \frac{u_{jet} \cdot l}{u_{max}D},
\end{equation}
\noindent where $u_{jet}$ denotes the velocity of the jet. In this study, $Q$ does not exceed 0.07, which is twice the value used by Chen et al. \citep{Chenetal2023}, owing to the double width of the jets compared to their study.

\subsection{Model-free DRL}\label{subsec2}
In contrast to supervised learning, which relies on paired data, and unsupervised learning, which is employed for clustering and dimensionality reduction, RL operates in a Markov decision process, learning through iterative interaction and feedback. The RL learning process is based on two primary components: the agent and the environment. These elements interact continuously, refining strategies through experiences across three essential channels: state ($s_t$), action ($a_t$), and reward ($r_t$). During each cycle $t$, the agent observes the current state $s_t$ of the environment, takes action $a_t$ that impacts the environment, and subsequently receives a new state $s_{t+1}$ along with a reward $r_t$. This reward serves as feedback guiding the agent to identify and select optimal actions. In the context of the AFC problem studied here, the environment for MFRL is simulated by DNS of the flow around a square cylinder confined between parallel walls, performed using Nek5000. Details about the agent will be provided in the following paragraphs.

There are several RL algorithms used for processing information, updating policies, and approximating value functions, including asynchronous advantage actor–critic \citep{Mnihetal2016}, trust region policy optimisation \citep{Schulmanetal2015}, and proximal policy optimisation (PPO) \citep{Schulmanetal2017}. Among these, PPO is particularly favoured for its straightforward implementation and stable performance. This stability is achieved through a clipped objective function, ensuring that updates to the policy are conservative and stay close to the previous policy. In this study, the episode-based PPO algorithm is employed, where the agent consists of an actor and a critic. The actor determines actions to take, while the critic evaluates these actions by estimating expected cumulative rewards. For a more detailed technical explanation of PPO, please refer to Schulman et al. \citep{Schulmanetal2015} and Yousif et al. \citep{Yousifetal2023d}.

\begin{figure}[htbp]
  \centerline{\includegraphics[scale=0.38]{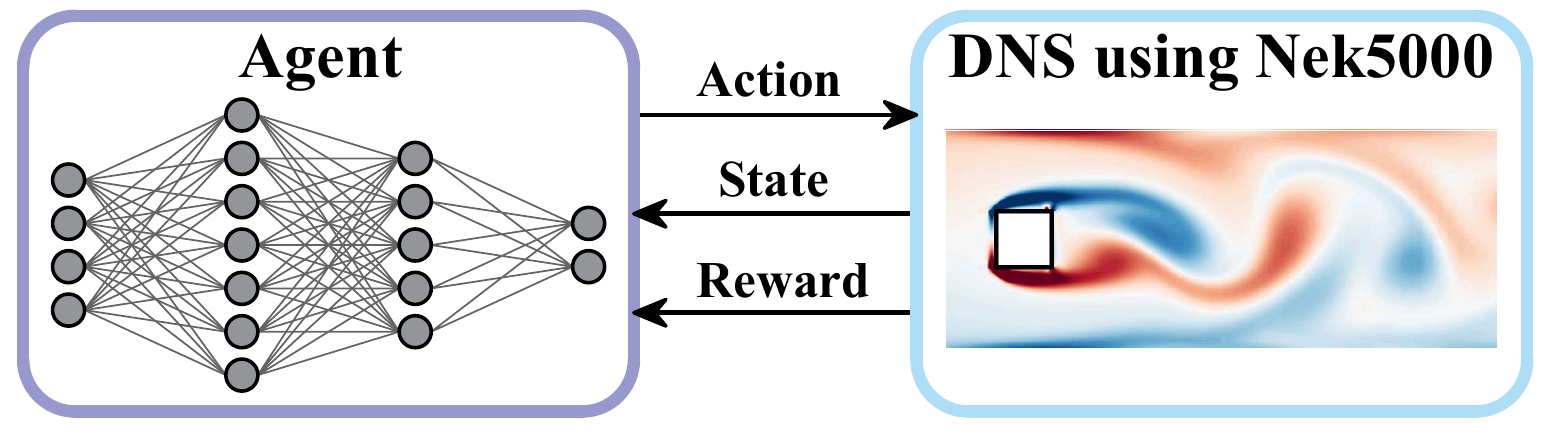}}
  \caption{Schematic of a typical DRL. The environment, represented by a DNS simulation of flow past a square cylinder using Nek5000, interacts with the deep neural network (DNN) agent through three essential channels: reward, state, and action.}
\label{fig:F2}
\end{figure}

As previously mentioned, DRL is recognised as an efficient method for actively controlling wake flow to reduce vortex shedding, leveraging the decision-making capability of RL combined with the mapping capabilities of DNNs, as depicted in Figure~\ref{fig:F2}. Following the approach of Rabault et al. \citep{Rabaultetal2019} and Li and Zhang \citep{Li&Zhang2022}, the DNNs used in this study are feed-forward neural networks comprising an input layer, two dense layers with 512 neurons each, and an output layer. The actor DNN takes $s_t$ as input and generates the policy $\pi_{\theta}(a_t \mid s_t)$, representing the probability distribution of $a_t$, where $\theta$ denotes the parameters of the actor. The process of optimising the policy involves updating $\theta$ to maximise the expected cumulative reward. The cumulative reward $R_t$ for the $t$-th cycle is calculated as $R_t = \sum_{t ^{\dagger}=t}^{T} \gamma ^{(t ^{\dagger}-t)}r_{t ^{\dagger}}$, where $T$ denotes the cycle at which the episode terminates, $\gamma$ is the discount factor, and $\gamma ^{(t ^{\dagger}-t)}$ is the $(t ^{\dagger}-t)$-th power of the discount factor. In this study, $T$ and $\gamma$ are set to 80 and 0.97, respectively.

\begin{figure}[htbp]
  \centerline{\includegraphics[scale=0.52]{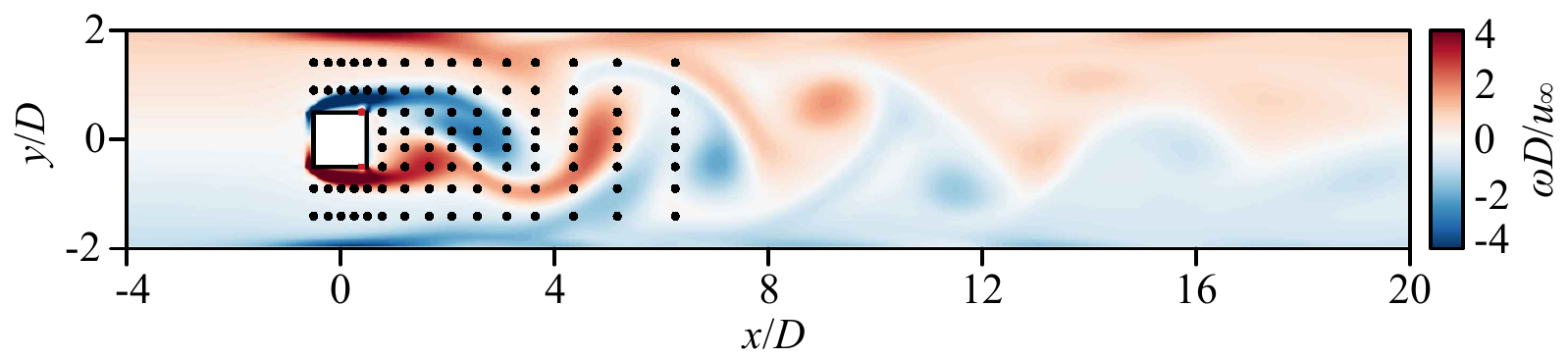}}
  \caption{Instantaneous vorticity in the wake behind the square cylinder immediately after flow initialisation without active control. The wake region of the square cylinder is monitored by 100 probes (marked with black dots), while the positions of the control jets are indicated by red dots.}
\label{fig:F3}
\end{figure}

Once a basic understanding of the agent and environment is established, the three interaction channels (state $s_t$, action $a_t$, and reward $r_t$) will be introduced. Specifically, the state is defined by partial observations of the velocity field collected by multiple probes strategically positioned around the cylinder and in the near-wake region, as illustrated in Figure~\ref{fig:F3}. In this study, 100 probes are used, a balance similar to the approach in Rabault et al. \citep{Rabaultetal2019} and Li and Zhang \citep{Li&Zhang2022}, to effectively capture flow characteristics without significantly increasing computational demands. During interactions, these probes provide streamwise and spanwise velocity components from the environment, representing flow features without affecting the original flow field. The actor determines an action based on the input state and applies it in the Nek5000 simulation environment. This action corresponds to adjusting the mass flow rate of the jets, as discussed in Section \ref{subsec1}. Each action modifies at intervals of $\Delta t = 0.2$ time units, equivalent to 7.77\% of a vortex shedding cycle. The primary objective of this study is to suppress vortex shedding in the wake flow confined behind the square cylinder, specifically aiming to minimise kinetic energy. Following the approach of Li and Zhang \citep{Li&Zhang2022}, the fluctuation of kinetic energy $S_{e}$ is used as an indicator of vortex shedding intensity, thereby defining the goal as decreasing $S_{e}$. The equation for $S_{e}$ is defined as:
\begin{equation} \label{eqn:eq6}
S_{e} = \sum_{i=1}^{n_{grid}} ((u_i - u_i^{SFD})^2 + (v_i - v_i^{SFD})^2),
\end{equation}
\noindent where $n_{grid}$ refers to the number of grid nodes in the Nek5000 simulation. $u_i$ and $v_i$ represent the streamwise and spanwise velocity components obtained by Nek5000 at the $i$-th grid node. The terms $u_i^{SFD}$ and $v_i^{SFD}$ denote the corresponding velocities derived using the selective frequency damping method \citep{Akerviketal2006}. Therefore, the reward $r$ is defined as the negative of the kinetic energy fluctuation, ensuring that maximising the reward corresponds to minimising the kinetic energy during the learning process. Each training episode spans $T_{max}= 16$ unit time, approximately 6.21 vortex shedding periods, or 3,200 simulation time steps. The PPO agent uses the adaptive moment estimation (Adam) optimiser \citep{Kingma&Ba2014} with a learning rate of $5 \times 10^{-5}$.

\subsection{Model-based DRL}\label{subsec3}
The previous subsection introduced the typical DRL framework, specifically model-free reinforcement learning (MFRL). In this subsection, the differences between MFRL and MBRL will be compared, and a detailed explanation of how the surrogate model integrates with the CFD environment will also be provided in detail. As stated in the introduction, the primary distinction between MFRL and MBRL is in whether the RL approach uses a dynamic surrogate model that closely mimics the real environment. In certain MBRL approaches, the RL agent is updated exclusively through interactions with the surrogate model. However, the efficacy of such MBRL methods heavily relies on the accuracy of the surrogate model. Inaccuracies can lead to erroneous policy decisions.

\begin{figure}[htbp]
  \centerline{\includegraphics[scale=0.45]{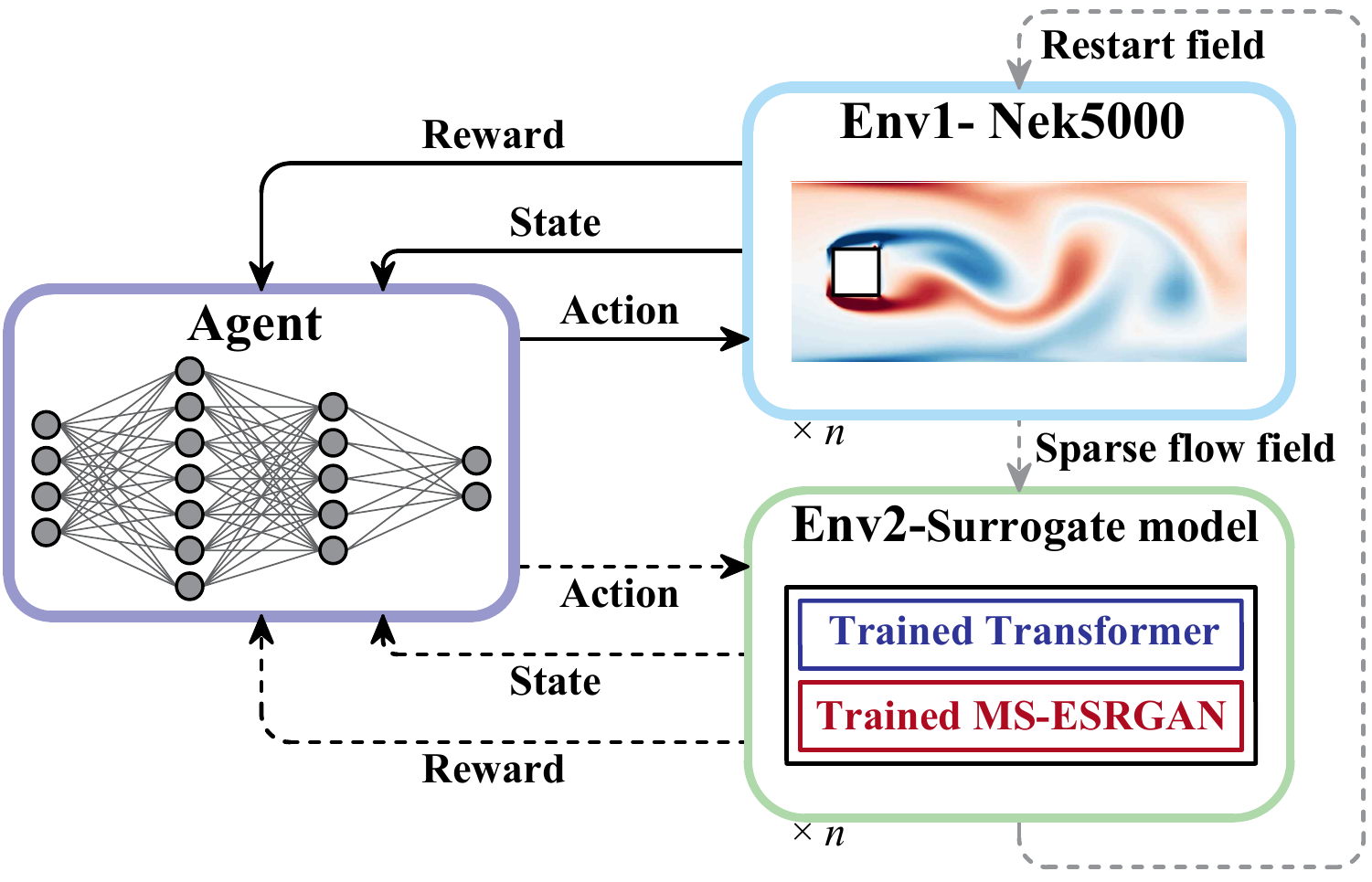}}
  \caption{Schematic of the proposed model-based DRL with alternating environments. DNS on Nek5000 (Env1) and deep-learning-based surrogate model (Env2) alternate every $n$ interactions with the DNN agent.}
\label{fig:F4}
\end{figure}

Therefore, the proposed DL–MBRL incorporates two alternating environments, as depicted in Figure~\ref{fig:F4}, to enhance the efficiency and accuracy of the RL process. The first environment, similar to that used in MFRL, involves DNS performed by Nek5000, depicted by the blue block in the figure. This environment is crucial because it generates highly accurate flow fields, ensuring that the RL agent learns and makes precise decisions. The second environment is a DL–SM, represented by the green block in the figure. This DL–SM integrates Transformer and MS–ESRGAN models, which will be elaborated upon in detail in Section \ref{subsec4}. This surrogate model is designed to considerably accelerate the training process with minimal error. In essence, the proposed DL–MBRL enhances the RL agent's optimisation by interacting with both a highly precise CFD environment and a fast, efficient surrogate model. The three channels facilitating communication between the agent and the environment, as well as the RL algorithms employed, remain consistent with those used in MFRL.

Specifically, the training of the DL–MBRL is divided into two stages: the CFD stage and the DL–SM stage, each consisting of $n$ interactions, where in this study, $n$ is specifically set to 5. \textbf{Initially}, the RL agent is initialised and begins interacting with the CFD environment. \textbf{During an interaction cycle of the CFD stage}, the CFD environment provides the agent with the current flow field information obtained from probes. Using this information, the agent selects a specific mass flow rate for the jets. Subsequently, the CFD environment applies this $Q$ to the flow field and returns the updated flow field along with the corresponding kinetic energy fluctuation. In each interaction in the CFD stage, the updated flow field information is sampled again from the probes and fed back to the agent. A sparse flow field is extracted from the simulation during each interaction and stored in the initial condition dataset for subsequent deep learning predictions in the DL–SM stage. \textbf{During an interaction cycle of the DL-SM stage}, the state remains the velocity components, but this state array is extracted from the sparse flow field predicted by the trained Transformer. It maintains the same format as in the CFD environment, ensuring consistency in the state representation. The spatial distribution of these data also aligns with the locations of the probes. The predicted sparse flow field updates the initial condition dataset used in the CFD stage, ensuring its readiness for subsequent interactions between the agent and DL–SM. This sparse flow field is then processed by the trained MS–ESRGAN for complete flow field reconstruction. The reconstructed flow field is used to calculate the kinetic energy fluctuation, which serves as the reward for the current interaction cycle. In the final step of the DL–SM stage, after $n$ interactions between the agent and the DL–SM, the flow field reconstructed by the MS–ESRGAN is prepared as the restart field for reconnecting with the CFD environment.

\subsection{Deep-learning-based surrogate model}\label{subsec4}
As outlined in Section \ref{subsec3}, the proposed DL–SM integrates two architectures: Transformer \citep{Vaswanietal2017} and MS–ESRGAN \citep{Yousifetal2021}. DL–SM employs a data-driven approach, requiring specific datasets for both submodels during training. The Transformer is trained using datasets comprising random $Q$ values for jets and their associated sparse flow fields, extracted from specific points in DNS flow fields. At time $t_m$, $Q_m$ modifies the sparse flow fields, resulting in updated sparse flow fields at $t_{m+1}$, a process iteratively applied until reaching the final time step $t_M$. To simulate the impact of $Q$ in the CFD environment, these sparse flow fields are added with their corresponding $Q$, creating augmented sparse flow fields. Subsequently, multiple successive time steps of these augmented fields serve as input for the Transformer model, which generates pure sparse flow fields as output. For the MS–ESRGAN model, the training dataset consists of pairs comprising sparse and high-resolution flow fields. Both datasets are partitioned into training and testing subsets, ensuring complete separation between the test and training data.

During the DRL interaction process, which also serves as the prediction phase for DL–SM, the Transformer takes as input a sequence of augmented sparse flow fields at time $[t_0,..., t_n]$. This sequence is the initial condition dataset (discussed in Section \ref{subsec3}), augmented with $Q$ values generated by the RL agent. The Transformer processes this sequence to predict the flow field at time $t_{n+1}$, which is then passed to MS–ESRGAN to generate a high-resolution flow field that aligns with the simulation data. In this study, the value of $n$ is fixed at 5, corresponding to the number of interactions discussed in the previous section. This iterative process involves advancing the input data by one time step at each iteration.

\begin{figure}[htbp]
  \centerline{\includegraphics[scale=0.4]{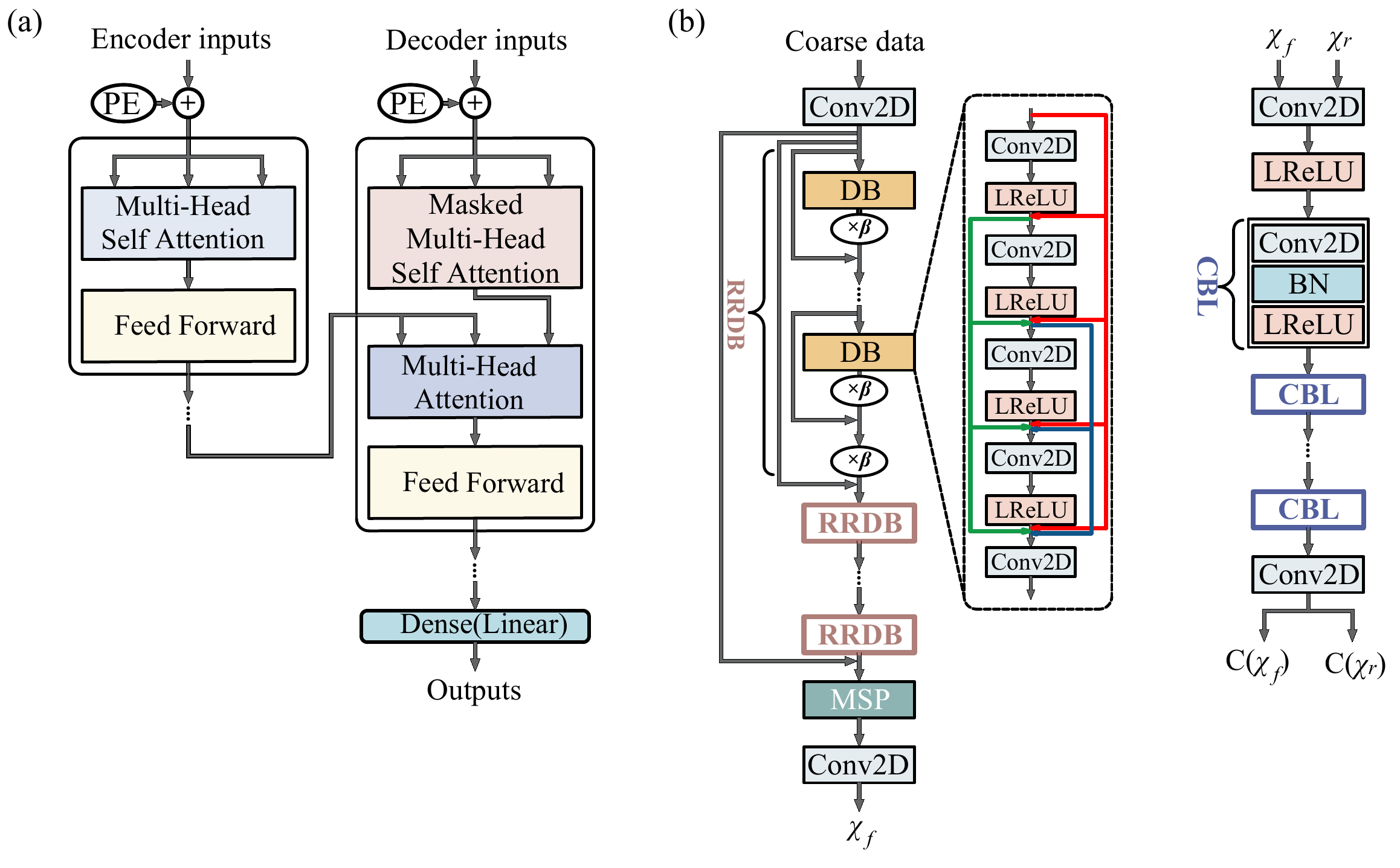}}
  \caption{Architecture of two deep learning models in the surrogate model: (a) Transformer and (b) MS–ESRGAN. Here, the left architecture of (b) is the generator, where $\beta$ is the residual scaling parameter set to 0.2 in this study, while the right architecture of (b) is the discriminator.}
\label{fig:F5}
\end{figure}

In the introduction, numerous studies have highlighted the potential of surrogate models in MBRL. However, these models have predominantly relied on neural ODEs and LSTM networks \citep{Hochreiter&Schmidhuber1997}. Despite their widespread adoption, these approaches are notably constrained by inherent limitations. Neural ODEs, while offering a continuous-time modelling framework, often face challenges related to computational efficiency and scalability. Similarly, LSTM networks, although powerful for sequential data processing, can encounter issues such as vanishing gradients and limited capability to capture long-term dependencies in highly complex datasets. Consequently, these models may not consistently deliver the required accuracy or performance in demanding applications, prompting interest in Transformers owing to their strong performance in handling long-range dependencies in sequences. Figure~\ref{fig:F5}(a) presents the architecture of the Transformer used in this study, featuring multiple encoder blocks (left block in Figuree~\ref{fig:F5}(a)) and decoder blocks (right block in Figure~\ref{fig:F5}(a)). Both encoder and decoder inputs are first processed by a positional encoder using sine and cosine functions, which incorporate the sequence information into the input vectors. Each encoder block comprises a multi-head self-attention layer followed by a feed-forward layer. The decoder blocks are analogous but incorporate an additional multi-head attention layer that pays attention to the encoder outputs. Furthermore, the multi–head self-attention layer in the decoder block is adapted to a masked multi–head self-attention layer to prevent information leakage. The multi–head self-attention mechanism primarily employs scaled dot–product attention to map an input query and its corresponding key-value pairs to an output. In contrast, the masked multi–head self-attention layer operates similarly but uses masked scaled dot–product attention to selectively attend to specific parts of the input sequence. The feed-forward layers consist of two dense layers with linear and rectified linear unit (ReLU) activation functions, transforming the input into a higher-dimensional space to capture relevant information before projecting it back to the original dimensionality. Additional details on the Transformer architecture are provided by Vaswani et al. \citep{Vaswanietal2017} and Yousif et al. \citep{Yousifetal2023c}. This study employs the squared L2 norm error as the loss function and uses the Adam optimiser for updating model weights.

MS–ESRGAN is a variant of the generative adversarial network, incorporating two adversarial neural networks: the generator and the discriminator. The generator's objective is to produce fake images that mimic the statistical characteristics of real images, while the discriminator tries to differentiate between fake and real images. Through effective training, the generator aims to generate images that are indistinguishable from real ones according to the discriminator's assessment. The MS–ESRGAN generator has a specific architecture that includes an initial convolutional layer, multiple deep residual in residual dense blocks, three parallel convolutional submodels with different kernel sizes termed multi–scale parts (MSP), and a final convolutional layer, illustrated in the left column of Figure~\ref{fig:F5}(b). The discriminator, shown in the right column of Figure~\ref{fig:F5}(b), consists of a series of convolutional layers, batch normalisation, and leaky ReLU layers. The competition between the generator and discriminator results in a combined loss function, which includes the adversarial loss of the generator and the mean-squared error (MSE) between generated and ground truth data. Further details are provided by Yousif et al. \citep{Yousifetal2021}. In this study, training data are divided into mini-batches of size 16, and the Adam optimisation algorithm is used to update model weights.

\section{Results and discussion}\label{sec3}
This section begins with the exploration of training strategies aimed at achieving low shedding energy, focusing on the convergence of RL algorithms in both standard MFRL and the proposed DL–MBRL approaches. Subsequently, the results from the DL–SM showcase the surrogate model's ability to simulate the real CFD environment accurately. The final subsection evaluates and compares the control performance of optimal strategies derived from both standard MFRL and the proposed DL–MBRL.

\subsection{Convergence of learning}\label{subsec5}
\begin{figure}[htbp]
  \centerline{\includegraphics[scale=0.32]{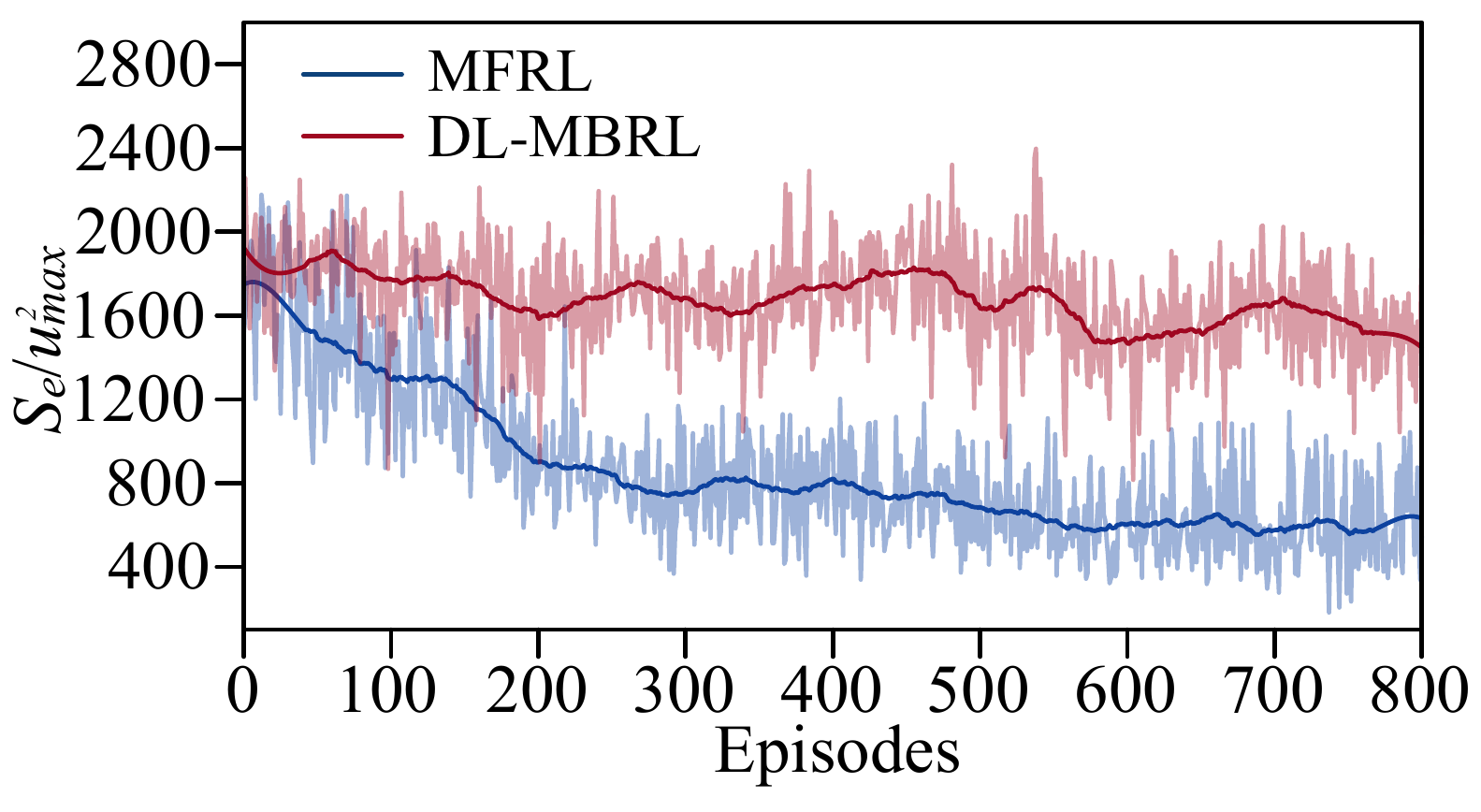}}
  \caption{Evolution of averaged shedding energy during the training process using MFRL and DL–MBRL. The light curves represent shedding energy, while the dark curves indicate smoothed shedding energy achieved through the Savitzky–Golay filter.}
\label{fig:F6}
\end{figure}

Following the methodologies outlined by Rabault et al. \citep{Rabaultetal2019} and Li and Zhang \citep{Li&Zhang2022}, a random reset mechanism is implemented in the training processes of both MFRL and DL–MBRL. Specifically, at the onset of each new episode, there is a 20\% probability that the agent begins from a fixed initial condition; otherwise, it continues from the final state of the previous episode. This random reset, coupled with exploration noise, introduces fluctuations in the averaged energy curves of MFRL and DL–MBRL during their training phases, as illustrated in Figure~\ref{fig:F6}. For enhanced clarity, the dark curves in the figure represent data smoothed using the Savitzky–Golay filter. The training process spans up to 800 episodes, with updates to agent network parameters occurring every 10 episodes. Importantly, there is a notable discrepancy in total training time between DL–MBRL and MFRL. Specifically, MFRL requires 41.87 h to reach the optimal strategy, whereas DL–MBRL accomplishes this in just 20.62 h, representing a 49.2$\%$ reduction in training time. Both training procedures were performed under identical conditions on the same machine to ensure a fair comparison.

In terms of convergence, MFRL stabilises at an average energy level of approximately 450 after 500 episodes, whereas DL–MBRL achieves an average energy level of 1,500. DL–MBRL exhibits a less favourable convergence trajectory compared to MFRL, partly attributed to the inherent randomness introduced by DL–SM during training. However, this poor convergence result does not imply poor control performance during testing. During testing, DL–SM is not used; its role is limited to simulating the real environment during training. Therefore, the randomness introduced by DL–SM influences only the training process and not the control performance during testing. The subsequent results detailed in Section \ref{subsec7} illustrate that despite DL–MBRL's training curve not converging as closely as MFRL's, the control performance in testing using the optimal strategy derived from DL–MBRL matches or even surpasses that of MFRL.

\subsection{Results of the deep-learning-based surrogate model}\label{subsec6}
\begin{figure}[htbp]
  \centerline{\includegraphics[scale=0.4]{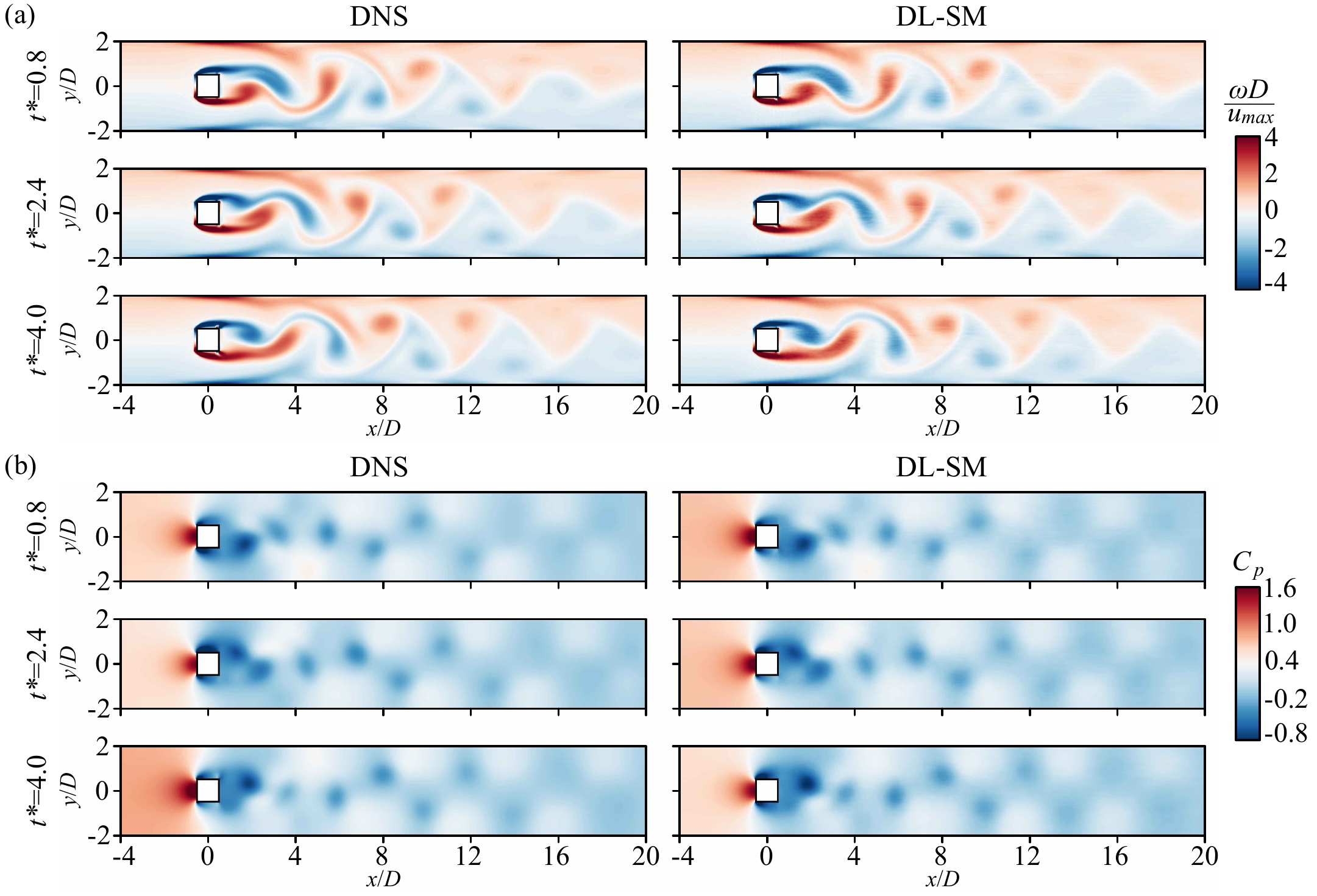}}
  \caption{Instantaneous (a) vorticity field and (b) pressure field of the flow past a square cylinder confined between parallel walls at three different time instants. Both reference (DNS) and predicted (DL–SM) data are shown.}
\label{fig:F7}
\end{figure}

In MBRL, the surrogate model plays a pivotal role in accurately predicting environmental dynamics, thereby influencing policy optimisation and training efficiency. Figure~\ref{fig:F7} illustrates the instantaneous non-dimensional vorticity $\omega D/u_{max}$ (a) and pressure $Cp$ (b) fields from DNS alongside predictions at three different non-dimensional time steps ($t^*=tu_{max}/D=$ 0.8, 2.4, 4.0). The figure shows that DL–SM can predict instantaneous flow patterns and pressure distributions with a commendable agreement with the DNS data across various time points, highlighting its capability to model complex fluid dynamics accurately. The pressure field predicted by DL–SM is used as one of the restart fields in conjunction with the CFD environment.

\begin{figure}[htbp]
  \centerline{\includegraphics[scale=0.42]{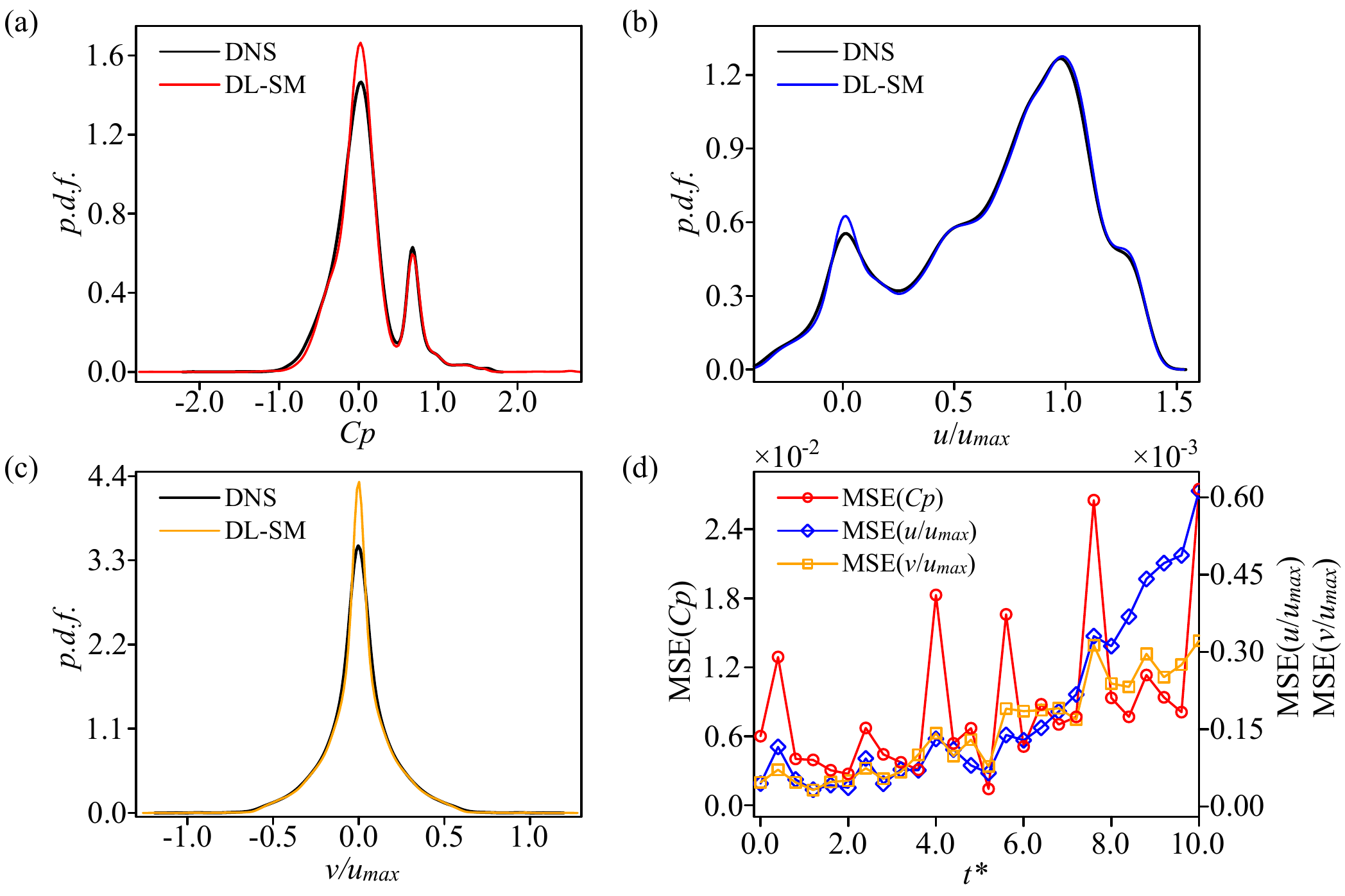}}
  \caption{(a) Probability density functions ($p.d.f.$) of the predicted pressure field. (b) $p.d.f.$ of the predicted streamwise velocity field. (c) $p.d.f.$ of the predicted spanwise velocity field. (d) Overall MSE of the flow fields ($Cp$, $u/u_{max}$, and $v/u_{max}$) predicted by DL–SM across non-dimensional time ($t^{*}$).}
\label{fig:F8}
\end{figure}

Figure~\ref{fig:F8}(a), (b), and (c) illustrate the probability density functions ($p.d.f.$) for the non-dimensional pressure $Cp$, streamwise velocity $u/u_{max}$, and spanwise velocity $v/u_{max}$, respectively. In each case, the DNS results (black lines) closely match the DL–SM results (coloured lines), demonstrating consistent agreement with minor deviations owing to the complexities and extreme values in certain regions. The DL–SM effectively captures the primary characteristics of these distributions, showcasing its proficiency in learning and representing essential flow dynamics features. Figure~\ref{fig:F8}(d) presents the overall MSE for $Cp$, $u/u_{max}$, and $v/u_{max}$ across time. Although the trends in MSE show a gradual increase over time, the DL–SM results remain consistent with DNS data. This increase in MSE is attributed to the accumulation of prediction errors at each time step. To address this, the surrogate model is swapped with the CFD environment every five interactions with the agent. This interval of five interactions was determined through testing to balance computational efficiency and accuracy. These qualitative and quantitative comparisons underscore the effectiveness of DL–SM as a computationally efficient substitute for the CFD environment, delivering high-fidelity approximations of fluid behaviour while significantly reducing computational costs.

\subsection{Control effect of MFRL and DL-MBRL}\label{subsec7}
\begin{figure}[htbp]
  \centerline{\includegraphics[scale=0.43]{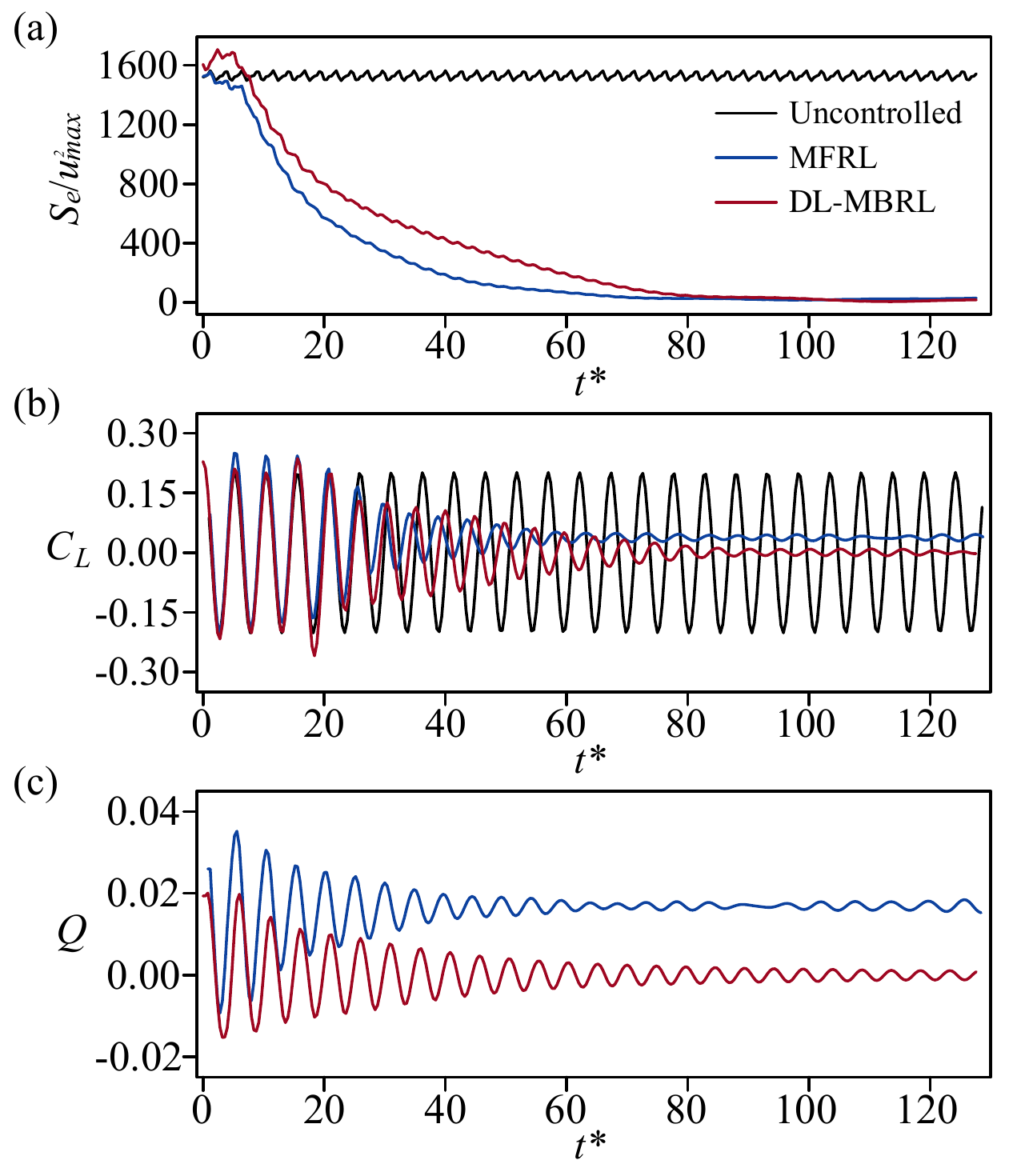}}
  \caption{Comparison of active control performance between DL–MBRL (red lines) and standard MFRL (blue lines) for flow past a square cylinder confined between parallel walls. Temporal variations of (a) shedding energy $S_{e}/u_{max}^2$, (b) lift coefficient $C_{L}$, and (c) mass flow rate $Q$. The black lines represent the uncontrolled case.}
\label{fig:F9}
\end{figure}

The effectiveness of trained MFRL and DL–MBRL in reducing shedding energy $S_{e}/u_{max}^2$ is evaluated and depicted in Figure~\ref{fig:F9}(a). The evaluation spans 127.6 non-dimensional time $t^*$, showing both transient and asymptotic dynamics. Both methods start with identical initial conditions, with control applied at $t^*$ = 0. In the figure, the uncontrolled scenario (black line) maintains a consistently high $S_{e}/u_{max}^2$ value of approximately 1,600 throughout the duration. In contrast to the training process curves, both MFRL (blue line) and DL–MBRL (red line) considerably decrease $S_{e}/u_{max}^2$ over time, although DL–MBRL achieves this reduction slightly slower than MFRL during testing. The slower reduction observed with DL–MBRL during testing does not diminish its overall advantages. As detailed in Section \ref{subsec5} and Table~\ref{tab:Table1}, DL–MBRL requires significantly less training time compared to MFRL, with training durations of 41.8667 h for MFRL and 20.6167 h for DL–MBRL, respectively. This reduced training time for DL–MBRL translates into substantial savings in computational resources and time, facilitating faster model iteration and optimisation. Therefore, despite DL–MBRL achieving a slower reduction in shedding energy during testing, its efficiency in training time makes it a valuable method for practical applications.

Figure~\ref{fig:F9}(b) illustrates the lift coefficient $C_{L}$ as a function of non-dimensional time $t^*$. The uncontrolled scenario (black line) exhibits persistent high-amplitude oscillations with a zero mean throughout the period. Both MFRL and DL–MBRL effectively dampen these oscillations over time. Once stabilised, the $C_{L}$ for DL–MBRL returns to a zero mean, similar to the uncontrolled case. In contrast, MFRL results in a nonzero mean lift coefficient owing to asymmetries in mass flow rate $Q$ around the zero y-axis, depicted in Figure~\ref{fig:F9}(c). Table~\ref{tab:Table1} offers a comprehensive comparison of control performance metrics between MFRL and DL–MBRL. Both methods show comparable results in terms of mean shedding energy and standard deviation of $C_L$. However, DL–MBRL achieves a mean $C_L$ of 0.0006, which is significantly closer to zero compared to the mean $C_L$ of 0.0379 obtained with MFRL. This observation aligns with Figure~\ref{fig:F9}(b), confirming the presence of a nonzero mean lift coefficient phenomenon in the MFRL case.

\begin{table}
  \begin{center}
\scalebox{1.0}{
\begin{tabular}{cccccccccc} \hline\hline
&Case~ ~&~~Training time~ ~&~~$\overline{S_{e}}/{u_{max}^2}$~ ~&~~$\sigma_{C_L}$~ ~&~~$\overline{C_L}$  \\ \hline
&MFRL~ ~&~~$41.8667h$~~&~~$22.9230$~~&~~$0.0046$~~&~~$0.0379$&   \\
&DL–MBRL~ ~&~~$20.6167h$~~&~~$21.8515$~~&~~$0.0061$~~&~~$0.0006$& \\ 
\hline\hline
\end{tabular}}
  \caption{Comparative analysis of training duration and control performance between MFRL and DL–MBRL. $\sigma$ denotes the standard deviation.}
  \label{tab:Table1}
  \end{center}
\end{table}

The occurrence of a nonzero mean lift coefficient in MFRL has been documented in several studies. For instance, Tang et al. \citep{Tangetal2020} observed that flow control using MFRL was not perfectly stable, displaying a nonzero mean and irregular jumps in the lift coefficient profile. They tried to address this issue by employing various interpolation functions for the mass flow rate value to minimise large deviations between consecutive $Q$ values. However, while the $C_{L}$ jumps were resolved, the nonzero mean $C_{L}$ remained. Similarly, Ren et al. \citep{Renetal2021} identified that the nonzero mean $C_{L}$ was due to the jet pair problem, where one side continuously blew while the other side consistently sucked. They adjusted the factor of the $C_{L}$ term in the reward function and discovered an optimal factor to mitigate these undesirable asymmetric lift fluctuations. These studies collectively indicate that the nonzero mean lift coefficient arises from insufficient exploration in RL, leading the agent to become trapped in a local optimum. Adequate exploration is crucial for enabling the agent to discover optimal strategies by trying diverse actions and avoiding local optima. Several methods can enhance exploration: 1. Applying epsilon-greedy strategies for the agent; 2. Modifying the reward function to encourage the agent to explore new or less-visited states, as implemented by Ren et al.; 3. Introducing randomness into the action selection process, unlike Tang et al.'s method, which smoothed action changes but failed to overcome the nonzero mean $C_{L}$. In this study, the proposed DL–MBRL inherently enhances exploration without the need for manual adjustments. By leveraging the DL–SM, DL–MBRL introduces variability and unpredictability into the environment, prompting the agent to continually seek new strategies. The inherent randomness of DL models, as discussed in Section \ref{subsec5}, improves the agent's ability to generalise across different situations. This effectively addresses the nonzero mean $C_{L}$ issue without relying on specific exploration techniques.

\begin{figure}[htbp]
  \centerline{\includegraphics[scale=0.4]{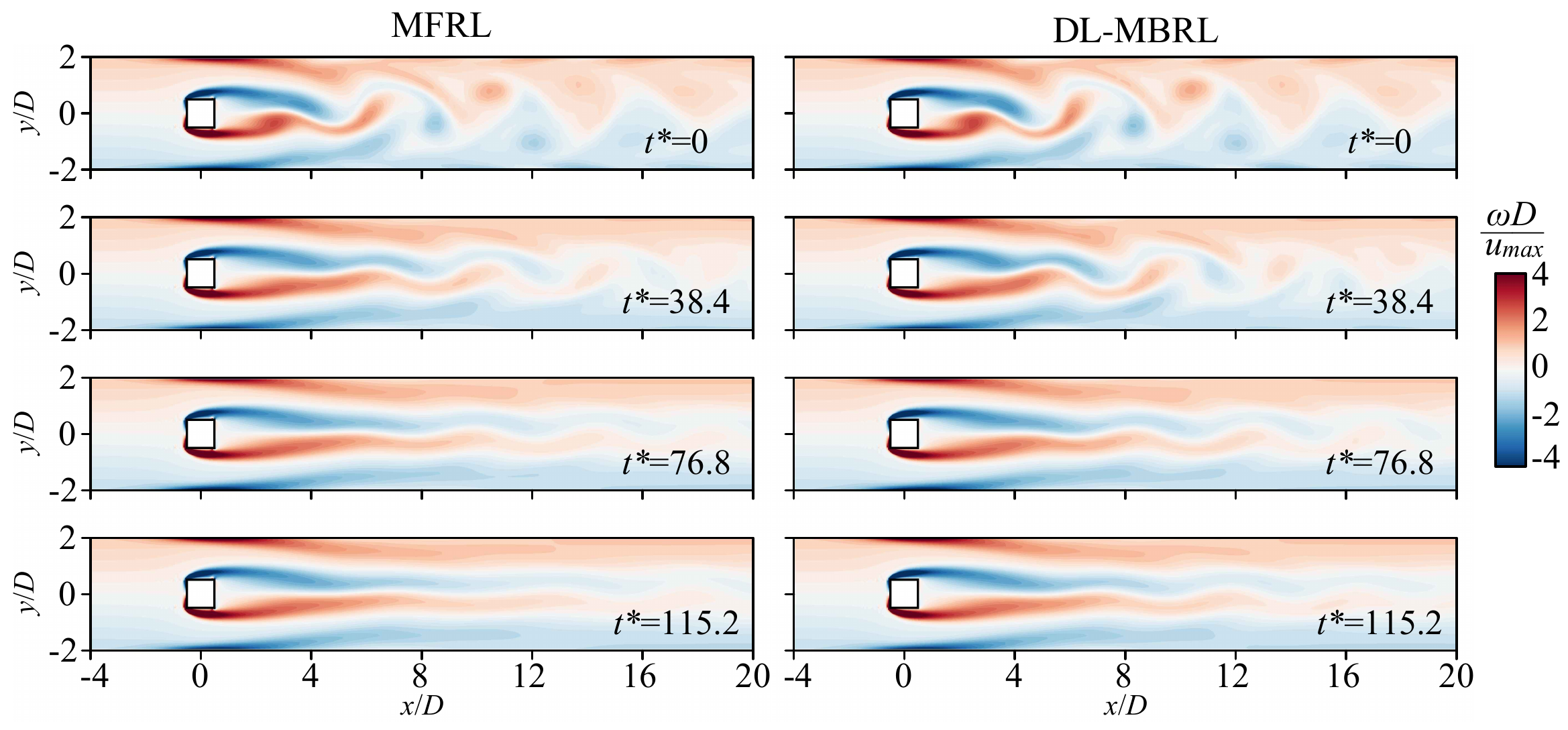}}
  \caption{Instantaneous vorticity fields of flow past the square cylinder confined between parallel walls are captured at four temporal instances, reflecting the results controlled by DL–MBRL (left column) and MFRL (right column).}
\label{fig:F10}
\end{figure}

Figure~\ref{fig:F10} presents the instantaneous vorticity contours at four temporal instances around the square cylinder, illustrating the stabilisation effects of MFRL and DL–MBRL. The first row depicts the initial state at $t^*=0$, showing the vorticity field without control. This uncontrolled flow provides a comparison of how the flow fields evolve under the MFRL and DL–MBRL control strategies. In the uncontrolled case, the flow regime behind the square cylinder is characterised by a Kármán vortex street, appearing as a regular and strong vortex train. As the control strategies are applied, both MFRL and DL–MBRL gradually reduce the intensity of vortex shedding, resulting in smoother flow structures. The vorticity contours reveal that the formation zone of the shedding vortices extends further downstream with the implementation of control. By the final temporal instance, both MFRL and DL–MBRL have significantly elongated the recirculation zone, indicating effective and sustained control. The contours demonstrate that DL–MBRL is equally effective as MFRL.

\begin{figure}[htbp]
  \centerline{\includegraphics[scale=0.4]{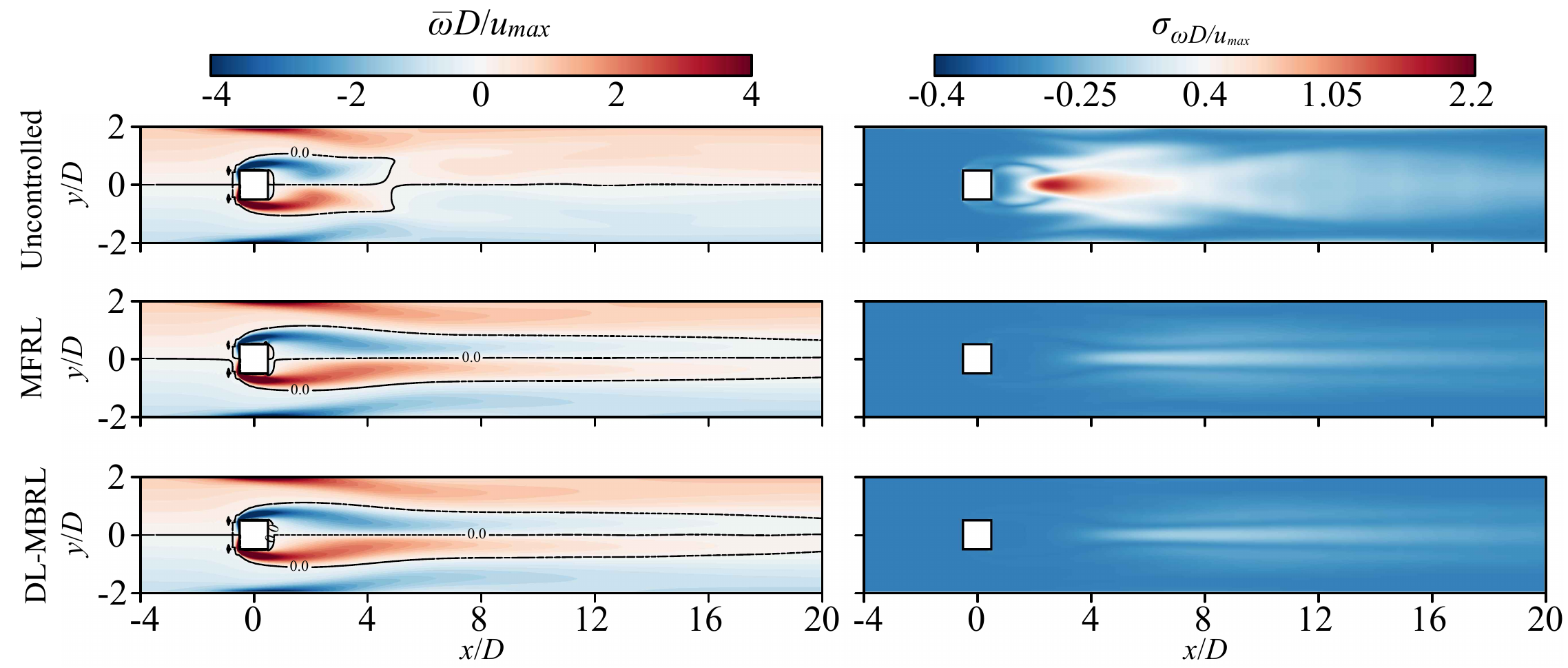}}
  \caption{Comparison of the wake flow morphology without control (upper row) and with control by MFRL (middle row) and DL–MBRL (lower row). The left column shows the mean vorticity field, with black dashed lines representing zero vorticity isolines, while the right column displays the standard deviation ($\sigma$) of the vorticity field. These statistics are averaged over many vortex shedding periods for the uncontrolled case, and the pseudo-periodic regime is considered for the MFRL and DL–MBRL cases.}
\label{fig:F11}
\end{figure}

To statistically evaluate the impact of the control strategies, Figure~\ref{fig:F11} provides a visual comparison of the time-averaged vorticity and the standard deviation of vorticity for uncontrolled, MFRL-controlled, and DL–MBRL-controlled flows. As observed in the instantaneous snapshots, the left column shows that both MFRL and DL–MBRL control strategies substantially enlarge the recirculation zone and reduce fluctuations, leading to a more streamlined recirculation zone. This indicates enhanced flow stability and energy efficiency. The right column depicts the standard deviation of vorticity, revealing that both MFRL and DL–MBRL suppress the large fluctuations and unsteady vortex shedding present in the uncontrolled wake flow. Notably, the DL–MBRL method further reduces the standard deviation of vorticity compared to MFRL, indicating a higher degree of flow stabilisation and control effectiveness.

\begin{figure}[htbp]
  \centerline{\includegraphics[scale=0.48]{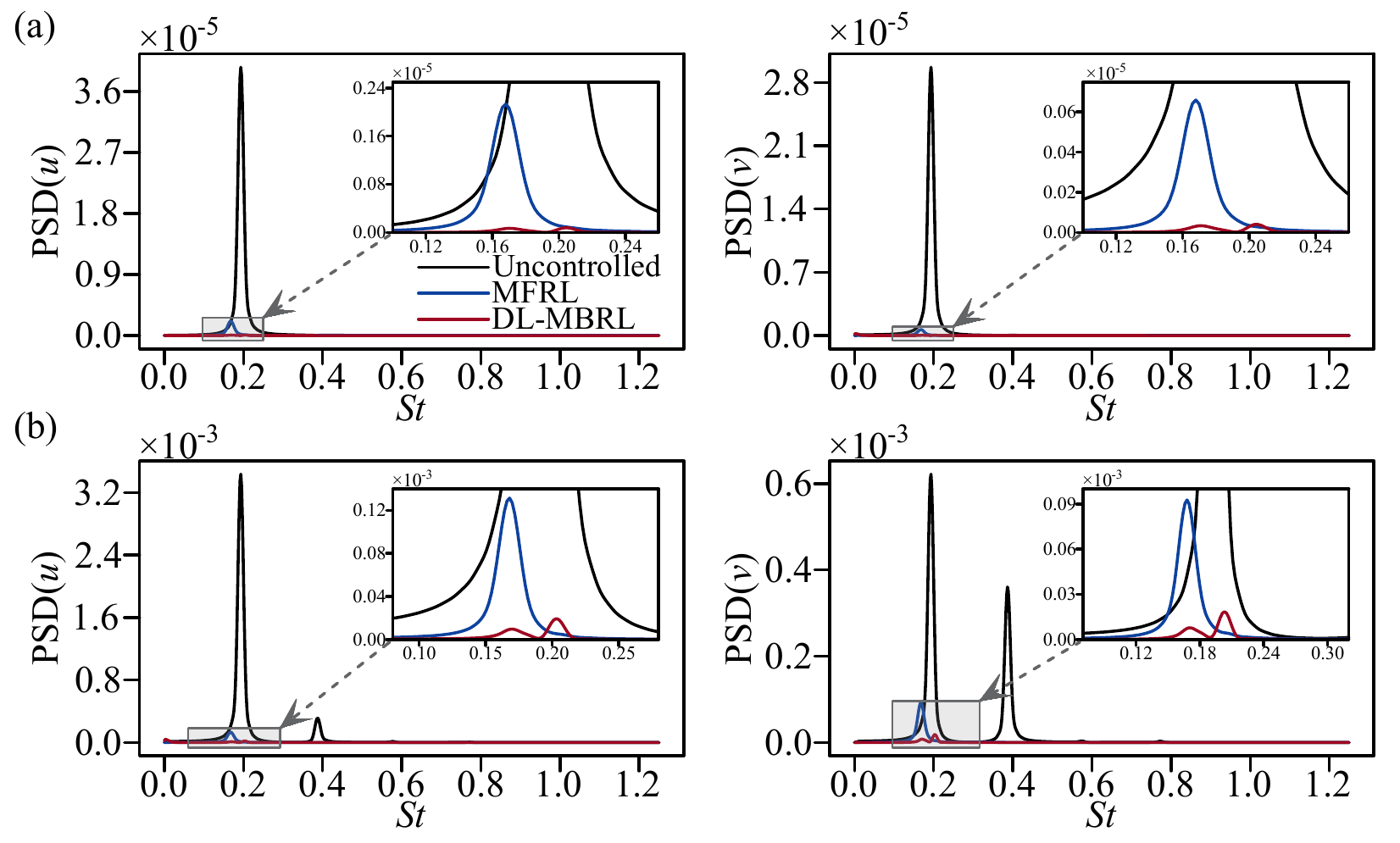}}
  \caption{Power spectrum density (PSD) of the streamwise (left) and spanwise (right) velocity fluctuations at two different locations: (a) ($x/D$, $y/D$)=(1, 1) and (b) ($x/D$, $y/D$)=(6, 1). The non-dimensional frequency is represented by the Strouhal number, $St = fD/u_{max}$.}
\label{fig:F12}
\end{figure}

Figure ~\ref{fig:F12} shows the PSD of the streamwise and spanwise velocity fluctuations at two different locations for uncontrolled and controlled cases, highlighting the oscillatory characteristics of the velocity components. In the uncontrolled case, the PSD exhibits a dominant non-dimensional frequency at $St \approx 0.2$, consistent with the Strouhal number derived from the vortex shedding frequency discussed in Section \ref{subsec1}. With the implementation of flow controls, the prominent peak in each subplot is either considerably attenuated or nearly vanished, indicating that the actions derived from the trained agents effectively suppress the vortex formation and shedding process. At the location ($x/D$, $y/D$)=(1, 1), the MFRL method reduces the PSD peak of streamwise velocity fluctuations by 95\% and the spanwise velocity fluctuations by 97\%, while the DL–MBRL method further reduces these PSD values by 99\%, effectively flattening the peaks and indicating a more stabilised flow with minimal fluctuations. Similar results at the location ($x/D$, $y/D$)=(6, 1) demonstrate that the DL–MBRL control strategy outperforms the MFRL approach in suppressing unsteady vortex shedding and stabilising the flow.

\section{Conclusions}\label{sec4}
This study proposes a DL–MBRL approach to actively control the 2D wake flow past a square cylinder confined between parallel no-slip walls using a pair of antiphase jets. The DL–MBRL approach consists of a DNN agent and two alternating environments: a deep learning-based surrogate model (DL–SM) and a CFD simulation. This method aims to suppress wake vortex shedding accurately while significantly reducing computational costs. The DL–SM, constructed using a Transformer and MS–ESRGAN, efficiently models complex nonlinear flow dynamics and emulates the CFD environment. The Transformer captures the temporal evolution of flow fields from spatially sparse data, simplifying RL training using low-dimensional probe-measured data as the state of the RL. Concurrently, MS–ESRGAN is trained to perform super-resolution reconstruction of the predicted sparse flow fields, creating restart fields to reconnect with the CFD simulation.

The DL–SM was first trained using wake flow data, and the Transformer and MS–ESRGAN were trained independently. The combined performance of the Transformer and MS–ESRGAN was then evaluated. The resulting instantaneous flow fields and probability density functions showed excellent agreement with the DNS results, validating the DL–SM as a computationally efficient substitute for the CFD environment. Error analysis revealed an increase in MSE over time, requiring the replacement of the DL–SM with the CFD environment every five interactions during the DL–MBRL training process.

During the learning process of DL–MBRL, the trained DL–SM alternated with the CFD environment, resulting in a 49.2\% reduction in training time compared to typical MFRL. While MFRL required 41.87 h, DL–MBRL only took 20.62 h. Despite exhibiting less robust convergence than MFRL, DL–MBRL significantly reduced computational and time costs, enabling faster model iteration and optimisation. The optimal control strategies obtained from both trained MFRL and DL–MBRL were subsequently tested in the CFD simulation, achieving approximately a 98\% reduction in shedding energy and a 95\% reduction in the standard deviation of $C_L$. However, typical MFRL exhibited a nonzero mean lift coefficient phenomenon owing to insufficient exploration in RL. In contrast, the proposed DL–MBRL naturally enhances exploration without the need for manual techniques, leveraging the inherent randomness of DL–SM. This randomness improved the agent's ability to generalise across different situations, effectively addressing the nonzero mean $C_{L}$ issue.

The performance of the proposed DL–MBRL was further evaluated using instantaneous vorticity fields at several temporal instances, as well as the mean and standard deviation of vorticity. Both the instantaneous and statistical results showed that MFRL and DL–MBRL elongated the recirculation zone and suppressed unsteady vortex shedding, demonstrating that DL–MBRL matches MFRL in effectiveness while requiring significantly less training time. Additionally, the power spectrum density of the streamwise and spanwise velocity fluctuations confirmed that DL–MBRL reduces and even flattens the PSD peaks, highlighting the superior performance of the DL–MBRL control strategy in efficiently stabilising the flow.

This study demonstrates that integrating DRL with a deep-learning-based surrogate model can effectively control the flow past a square cylinder confined between parallel no-slip walls. The methodology leverages the DL–SM for efficient approximation and integrates CFD simulation for high accuracy, achieving stabilisation of the wake flow while significantly reducing computational costs. This approach shows potential for extending AFC to three-dimensional flow fields, which will require more sophisticated DRL models and more powerful surrogate models.

\begin{acknowledgments}
This work was supported by 'Human Resources Program in Energy Technology' of the Korea Institute of Energy Technology Evaluation and Planning (KETEP), granted financial resource from the Ministry of Trade, Industry \& Energy, Republic of Korea (no. 20214000000140).This work was supported by the National Research Foundation of Korea (NRF) grant funded by the Korea government (MSIT) (No. RS-2024-00406152).
\end{acknowledgments}


%


\end{document}